\documentclass[preprint,groupedaddress,superscriptaddress,showpacs,showkeys]{revtex4}

\usepackage{graphicx}
\usepackage{dcolumn}
\usepackage{bm}

\begin{document}

\title {Exact solutions of the sextic oscillator from the bi-confluent Heun equation}

\author{G\'eza L\'evai}
\affiliation{
             Institute for Nuclear Research, Hungarian Academy of Sciences (MTA Atomki),
             Debrecen, Pf. 51, 4001 Hungary;}
\email{levai@atomki.mta.hu}

\author{Artur M. Ishkhanyan}
\affiliation{Russian-Armenian University, Yerevan, 0051 Armenia;}
\affiliation{Institute for Physical Research, NAS of Armenia, 0203 Ashtarak, Armenia;}
\affiliation{Institute of Physics and Technology, National Research Tomsk Polytechnic University, Tomsk 634050, Russia;}
\email{aishkhanyan@gmail.com}

\begin{abstract}
The sextic oscillator is discussed as a potential obtained from the
bi-confluent Heun equation after a suitable variable transformation.
Following earlier results, the solutions of this differential equation
are expressed as a series expansion of Hermite functions with shifted
and scaled arguments. The expansion coefficients are obtained from a
three-term recurrence relation. It is shown that this construction
leads to the known quasi-exactly solvable form of the
sextic oscillator when some parameters are chosen in a
specific way. By forcing the termination of the
recurrence relation, the Hermite functions turn into Hermite polynomials
with shifted arguments, and, at the same time, a polynomial expression
is obtained for one of the parameters, the roots of which
supply the energy eigenvalues. With the $\delta=0$ choice the quartic
potential term is cancelled, leading to the {\it reduced} sextic
oscillator. It was found that the expressions for the energy eigenvalues
and the corresponding wave functions of this potential agree with those
obtained from the quasi-exactly solvable formalism. Possible
generalizations of the method are also presented.
\end{abstract}



\maketitle

\section{Introduction}
\label{intro}
Exactly solvable models have served as indispensable tools in the
exploration of the subatomic world since the dawn of quantum
mechanics. After the beginning of the computer era, the solution of
the quantum mechanical wave equations, e.g. the Schr\"odinger
equation could be performed with high accuracy using various numerical
techniques, nevertheless, exactly solvable models were still employed
widely as the starting point of numerical methods. Furthermore, these
models are often connected with various symmetries and algebraic
structures, so in addition to their practical use, their mathematical
beauty also still make them appealing.

In practical situations exact solvability means that the energy eigenvalues
and bound-state wave functions, as well as quantities related to scattering
(if applicable) can be expressed in closed analytical form. In most cases
this is achieved by transforming the Schr\"odinger equation into the
differential equation of some special function $F(z)$ of mathematical physics.
With this, the solution of the Schr\"odinger equation is
expressed as $\psi(x)=f(x)F(z(x))$, where the $z(x)$ function represents a
variable transformation. The choice of $F(z)$ corresponds to identifying
various classes of solvable potentials. When $F(z)$ is the hypergeometric,
or confluent hypergeometric function, then the related exactly solvable
potentials are referred to as Natanzon \cite{natanzon} or Natanzon confluent
\cite{natconf} potentials that depend on six parameters. In practical
applications the bound-state wave functions reduce to Jacobi and generalized
Laguerre polynomials \cite{AS70}. Many of the most well-known textbook
examples (harmonic oscillator, Coulomb, P\"oschl--Teller, Scarf, Rosen--Morse,
etc.) appear as two- or three-parameter members of the shape-invariant
\cite{gendenshtein} subclass of the Natanzon (confluent) potential class.
The structure (shape) of these potentials is invariant under a transformation
of supersymmetric quantum mechanics (SSQM, for a review, see e.g. Ref.
\cite{ssqm}), hence their name.

More recently, a wider class of potentials began to attract  much attention.
In this case the Schr\"odinger equation is transformed into various versions
of the Heun equation \cite{heunRon,heunSlav,heunOlver}. These potentials depend
on more parameters, making them more flexible and versatile
\cite{LB,Batic1,SE-GHE,Discret, Batic2,SE-CHE,Ish17}. However, since the
solutions of the Heun equation are much less known than the (confluent)
hypergeometric functions, the exact analytic treatment of these potentials is usually
much more complicated technically than that of the Natanzon (confluent)
potentials. In certain cases the $F(z)$ function is expressed as an expansion
of other special functions \cite{Ish17,Svart,Erd,Schmidt,EJFigueiredo}. In
Ref. \cite{Ish17}, for example, a systematic discussion of
potentials obtained from the bi-confluent Heun equation (BHE) is presented,
with bound-state wavefunctions expressed as an expansion in terms of Hermite
functions. Other examples involving truncated series solutions of the Heun
equations in terms of functions of the hypergeometric class are presented in
Refs. \cite{Ortega1,Ish_InvSqrt,Ish_LambertWsing,Ish_CES_gen,Ortega2,Ish_Third}

The concept of solvability has also been extended, revealing further aspects
of quantum mechanical potentials problems. In the case of conditionally
exactly solvable (CES) potentials, exact solutions are obtained only for
potentials in which some of the potential parameters are correlated, or are
restricted to constant values.
This concept was introduced in Ref. \cite{dutra} in relation with
potentials discussed earlier \cite{stillinger}. However, due to some
mathematical inconsistencies pointed out in Ref. \cite{znojil00}, the
first unquestionable example for conditionally exactly solvability was
the Dutt--Khare--Varshni
(DKV) potential \cite{dkv}, which has terms with fixed parameters. This
potential was later identified \cite{jmp01} as a member of the Natanzon-class,
and it was shown that the fixed potential parameters result from the
Schwartzian derivative that originates from the variable transformation
$z(x)$ (see e.g. \cite{jpa12,ijtp15} for the details and further similar
potentials). Another type of CES potentials was obtained as non-trivial
supersymetric partners of shape-invariant potentials \cite{JR97}, which
are beyond the Natanzon class.

Quasi-exactly solvable (QES) potentials \cite{Tur86,Tur87,qes} typically
support infinite
number of bound states, of which, however, only the lowest few are discussed
exactly. In this case the $F(z)$ function appearing in the bound-state
eigenfunctions is expressed as a power series expansion, with coefficients
satisfying a three-term recursion relation. For some potential parameters the
infinite series can be terminated, and $F(z)$ reduces to a polynomial.
Perhaps the most well-known example for QES potentials is the sextic oscillator
defined in one dimension or as a radial potential \cite{qes}. However, its
solutions cannot be obtained for arbitrary potential parameters, only for
certain correlated parameter sets. This potential has been applied in realistic
calculations describing, for example, shape phase transitions of various nuclei
\cite{prc04,prc10,Bud16}.

Due to the various concepts of solvability, sometimes the same potentials can
be discussed within different frameworks. Here we report on a study of this
kind: the sextic oscillator, which has been described as a QES potential is
analyzed in terms of the BHE approach \cite{Ish17}, and the results from
the two methods are compared.
In the QES approach the parameters of the sextic oscillator are correlated,
while in Ref.\cite{Ish17} this question is not discussed, so the question
whether the two methods describe the same potential or not, is raised
naturally.

In Sec. \ref{qes-sum} the sextic oscillator is revieweded  as a QES potential.
The formalism of the BHE approach is outlined in Sec. \ref{bhe-sum}, while in
Sec. \ref{comparison} the method is used to generate the bound-state solutions
and the corresponding energy eigenvalues of the reduced sextic oscillator, i.e.
the case containing sextic, quadratic and inverse quadratic terms, but not the
quartic one. Finally, the results are summarized in Sec. \ref{summary}, together
with further possible applications of the formalism.

\section{The sextic oscillator as a QES potential}
\label{qes-sum}

The most general form of the sextic oscillator is
\begin{equation}
V(x)=V_{-2}x^{-2}+V_2 x^2 + V_4 x^4 + V_6 x^6\ .
\label{sextosc}
\end{equation}
It is assumed that $V_6>0$, so the potential tends to plus infinity,
meaning that it has infinitely many bound-state solutions. For $V_6=V_4=0$,
(\ref{sextosc}) reduces to the radial harmonic oscillator.
This potential can be interpreted as the radial component ($x\in[0,\infty)$) of
a spherically symmetric potential in $D$ dimension. Here $V_{-2}$ depends on
$D$ and on the orbital angular momentum $l$. Setting $V_{-2}=0$ the domain of
definition of (\ref{sextosc}) can be
extended to the full $x$ axis, $x\in(-\infty,\infty)$. In this case the bound-state
wave functions have definite parity, and the odd solutions that vanish at $x=0$
correspond to the solutions of the radial problem.

Numerous attempts have been made to determine the bound-state energy eigenvalues
and the corresponding wave functions of (\ref{sextosc}) using various methods
including continued fractions \cite{Sin78}, $1/N$ expansion \cite{Dut88},
perturbation methods
\cite{Zno90,Wen96}, asymptotic iteration method \cite{Sou06}, etc.
It was found that exact analytic solutions are obtained
only for specific values of the potential parameters. Even then, only the
lowest few of the infinite number of energy eigenstates were obtained
analytically. The sextic oscillator
has thus been identified as a quasi-exactly solvable (QES) potential
\cite{Tur87,qes}.
Its usual parametrization is \cite{qes}
\begin{equation}
V(x)=\left(2s-\frac{1}{2}\right)\left(2s-\frac{3}{2}\right)x^{-2}
+[b^2-2a(2s+1+2M)]x^2 +2ab x^4 +a^2 x^6\ ,
\label{qes-sextosc}
\end{equation}
i.e. the four parameters $V_i$ ($i=-2$, 2, 4, 6) are correlated and reduce
to three real parameters $a$, $b$, $s$ and a non-negative integer $M$ that
determines the number $M+1$ of analytic solutions.

The solutions are written as
\begin{equation}
\psi(x)=(x^2)^{s-1/4}\exp\left(-\frac{ax^4}{4}-\frac{bx^2}{2}\right)F(x^2)\ .
\label{bswf}
\end{equation}
For $s=1/4$ and $s=3/4$ the centrifugal term in (\ref{qes-sextosc}) vanishes,
and (\ref{bswf}) reduces to the even and odd solutions of the one-dimensional
problem, respectively.
It has to be noted that polynomial potentials with higher order have
also been investigated using solutions containing the exponential form of
polynomials of $x^2$ \cite{Mag80}. It was found that this approach is suitable
for potentials with leading terms of the type $x^{4\nu+2}$. In this context the
unique role of the harmonic oscillator ($\nu=0$) and that of the sextic
oscillator ($\nu=1$) seems natural, and this finding also explains why the
quartic oscillator does not belong to the group of QES problems.

If $F(x^2)$ is an $M$'th order polynomial, then a second-order differential
equation  can be obtained from the Schr\"odinger equation after the separation
of the remaining factors of (\ref{bswf}). Matching the appropriate powers on
the two sides of the resulting equation, the coefficients of the polynomial
can be considered as the components of an $(M+1)$-dimensional vector that
satisfy an $(M+1)$-dimensional spectral matrix equation \cite{qes}. In a more
general approach, $F(x^2)$ can be expressed as a power series in $x^2$, and in
that case an infinite matrix is obtained. However, with the parametrization
used in (\ref{qes-sextosc}) it is possible to cancel certain off-diagonal matrix
elements of the infinite matrix, and thus to separate the $(M+1)$-dimensional
submatrix appearing in the upper left corner, leading to a quasi-exactly
solvable problem.

A more easily tractable problem appears for $b=0$, which cancels the quartic
component of (\ref{qes-sextosc}). In this {\it reduced} case of the sextic
oscillator the (unnormalized) wave functions and the
corresponding energy eigenvalues can be expressed in a relatively simple form
for the first few values of $M$ \cite{qes}:

$M=0$ In this case the ground-state energy is obtained with $n=0$, and
$E_0=0$, while the corresponding wave function is
\begin{equation}
\psi_0(x)=(x^2)^{s-1/4}\exp\left(-\frac{ax^4}{4}\right)\ .
\label{m0wf}
\end{equation}

$M=1$ Then the ground- and the first excited states are described with
$n=0$ and 1:
\begin{equation}
E_n=(-1)^{n+1}(32as)^{1/2}\ ,
\label{m1e}
\end{equation}
\begin{equation}
\psi_n(x)=(x^2)^{s-1/4}\exp\left(-\frac{ax^4}{4}\right)
 \left(ax^2-\frac{E_n}{4}\right)\ .
\label{m1wf}
\end{equation}
$\psi_0(x)$ has no node for $x>0$, while $\psi_1(x)$ has one.

$M=2$ Here the lowest three states are obtained with $n=0$, 1 and 2:
\begin{equation}
E_n=(n-1)[32a(4s+1)]^{1/2}\ ,
\label{m2e}
\end{equation}
\begin{equation}
\psi_n(x)=(x^2)^{s-1/4}\exp\left(-\frac{ax^4}{4}\right)
 \left(ax^4-\frac{E_n}{4} x^2 +\frac{E_n^2}{32a}-2s-1\right)\ .
\label{m2wf}
\end{equation}
The quadratic polynomial of $x^2$ in (\ref{m2wf}) has $n$ nodes
for $x^2>0$, reproducing the expected structure of the the
wave functions $\psi_n(x)$

$M=3$ The expressions for the first four states are somewhat more
involved in this case with $n=0$, 1, 2 and 3:
\begin{equation}
E_n=(-1)^{[\frac{n}{2}]+1}(32a)^{1/2}\left[
5(s+\frac{1}{2})+(-1)^{[\frac{n+1}{2}]}
\left(25(s+\frac{1}{2})^2-9s(s+1)\right)^{1/2}
\right]^{1/2}\ ,
\label{m3e}
\end{equation}
\begin{eqnarray}
\psi_n(x)&=&(x^2)^{s-1/4}\exp\left(-\frac{ax^4}{4}\right)
\nonumber\\
 &&\times\left(a^2x^6-a\frac{E_n}{4} x^4 +\frac{E_n^2-96a(s+1)}{32} x^2
 -\frac{E_n^3}{384a}+(7s+5)\frac{E_n}{12}\right)\ .
\nonumber\\
\label{m3wf}
\end{eqnarray}
The number of nodes of (\ref{m3wf}) appearing for $x>0$ is again $n$, as
expected.

Note that for the sake of simplicity, the wave functions $\psi_n(x)$ appear
here in an unnormalized
form \cite{qes}, however, the normalization constants can be determined in a
straightforward way. Note also that the wave functions vanish at $x=0$
as long as $s\ge 1/4$.
The properties of the polynomials appearing in these solutions have been
discussed in detail in Ref. \cite{Ben96},
with special attention to their comparison with other orthogonal polynomials.

\section{Potentials solvable in terms of the bi-confluent Heun equation}
\label{bhe-sum}

There has been increased interest in the bi-confluent Heun equation
\cite{heunRon,heunSlav,heunOlver} recently \cite{Batic1,Ish_InvSqrt,Ish_CES_gen,
F.M.andB.M.,HamzaviRajabi,Karw,Caruso,FonsecaBakke}, as a
second-order differential equation into which the stationary Schr\"odinger
equation can be transformed. In this Section we combine two approaches that
have been applied previously to obtain potentials solvable in terms of certain
special functions of mathematical physics, and specify them for the case of
the bi-confluent Heun equation. First we apply a variable transformation to
identify potentials solved in terms of the bi-confluent Heun equation, then
we review the method of expanding the solutions in terms of Hermite
functions \cite{Ish17}.

The method based on variable transformations originates from
Refs. \cite{BS62,Bose63}, and it resulted in the systematic classification of
shape-invariant potentials \cite{jpa89}. Its generalization has also been
used to give a systematic description of the wider Natanzon \cite{natanzon}
and Natanzon confluent \cite{natconf} potentials, i.e potentials with
bound-state solutions written in terms of a single hypergeometric or confluent
hypergeometric function. Essentially the same method was used in
Refs. \cite{LB,Discret,Ish17} to identify potentials solved in terms of the
bi-confluent Heun function. The method works well for other problems too.
For instance, it was applied to construct a variety of analytically solvable
quantum two-state models \cite{Shore}, in particular, via reduction of the
time-dependent Schr\"odinger equations to the confluent-hypergeometric and
bi-confluent Heun equations \cite{TL_BHE,Shahverd}.

\subsection{The potentials}
\label{bhe-sum-1}

The bi-confluent differential equation can be parametrized in
terms of five parameters as
\begin{equation}\frac{{\rm d}^2 u}{{\rm d}z^2}+\left(\frac{\gamma}{z}
+\delta +\varepsilon z\right)\frac{{\rm d} u}{{\rm d}z}
+\frac{\alpha z-q}{z} u\ .
\label{bhe}
\end{equation}
For $\alpha=0$ and $\varepsilon=0$ it essentially reduces to the
confluent hypergeometric differential equation \cite{AS70} .

In order to transform Eq. (\ref{bhe}) into the stationary Schr\"odinger
equation with units of $2m=\hbar=1$
\begin{equation}
\frac{{\rm d}^2\psi}{{\rm d}x^2}=[V(x)-E_n]\psi(x)
\label{sch}
\end{equation}
one may follow the procedure outlined in Ref. \cite{jpa89} and factorize
the wave function as $\psi(x)=f(x)u(z(x))$. Substiting this function in
(\ref{sch}) and comparing the result with (\ref{bhe}), one obtains
\begin{eqnarray}
E-V(x)&=&\frac{z'''}{2z'}-\frac{3}{4}\left(\frac{z''}{z'}\right)^2
+(z'(x))^2\left[\left(\frac{\gamma}{2}-\frac{\gamma^2}{4}\right)z^{-2}(x)
-\left(q+\frac{\gamma\delta}{2}\right)z^{-1}(x)\right.
\nonumber\\
&&\left. +\left(\alpha-\frac{\varepsilon}{2}-\frac{\delta^2}{4}
-\frac{\gamma\delta}{2}
\right)-\frac{\delta\varepsilon}{2}z(x) -\frac{\varepsilon^2}{4}z^2(x)\right]\ .
\label{emv}
\end{eqnarray}
Furthermore, according to Ref. \cite{jpa89}, the $f(x)$ function can be
expressed as
\begin{equation}
f(x)\sim (z'(x))^{-1/2}(z(x))^{\frac{\gamma}{2}}
\exp\left(\frac{\delta}{2}z(x)+\frac{\varepsilon}{4}z^2(x)\right)\ .
\label{fx}
\end{equation}

In the next step $z(x)$ is determined from the requirement that in
(\ref{emv}) the constant $E$ on the left side has to originate from certain
terms appearing on the right side in the form $(z')^2 z^k$, where $k=-2$, $-1$,
0, 1 and 1. Applying this requirement to potentials
related to the classical orthogonal polynomials resulted in the classification
of shape-invariant potentials \cite{jpa89}.
The method presented in Ref. \cite{Ish17} is similar in that individual
terms of the right-hand side of (\ref{emv}) correspond to $E$ on the left-hand
side, resulting in the differential equation
\begin{equation}
\frac{{\rm d}z}{{\rm d}x}=z^{m}/\sigma\ ,
\label{zdiff}
\end{equation}
where $m=-k/2$ and $\sigma$ is a technical parameter to be specified later.
With this choice, the five terms appearing in the square bracket in
(\ref{emv}) yield a constant term, provided that $m$ takes on the values
$-1$, $-1/2$, 0, 1/2 and 1. At the same time, the remaining terms on the
right-hand side of (\ref{emv}), the Schwartzian derivative, lead to a term
proportional to $z^{2m-2}$. In general, with the choice (\ref{zdiff}),
Eq. (\ref{emv}) turns into
\begin{eqnarray}
E-V(x)&=&\sigma^{-2}\left[-\frac{\gamma-m}{2}
\left(\frac{\gamma+m}{2}-1\right)z^{2m-2} (x)
-\left(q+\frac{\gamma\delta}{2}\right)z^{2m-1}(x)\right.
\nonumber\\
&&\left.+\left(\alpha-\frac{\varepsilon}{2}-\frac{\delta^2}{4}
-\frac{\gamma\varepsilon}{2}\right)z^{2m}(x)
-\frac{\delta\varepsilon}{2}z^{2m+1}(x) -\frac{\varepsilon^2}{4}z^{2m+2}(x)\right]
\ .
\nonumber\\
\label{emv2}
\end{eqnarray}
The five choices of $m$ result in the five different potentials listed in
Refs. \cite{LB,Discret,Ish17}. It is notable that the general discussion of
potentials related to the bi-confluent Heun equation in Ref. \cite{Batic1}
contain implicitly the same potentials, with the exception of the
$m=-1$ case.

The complete solution requires also solving the
differential equation (\ref{zdiff}). It is found that for $m=1$ one gets
$z(x)=\exp((x+x_0)/\sigma)$, while for the remaining cases
$z(x)=[(1-m)(x+x_0)/\sigma]^{1/(1-m)}$, where the coordinate shift $x_0$ is
an integration constant that can be omitted in most cases without the loss
of generality. The $\sigma$ parameter is also inessential, as it simply scales
the energy. It is also straightforward to see that the terms originating from
the Schwartzian derivative contribute to the constant term for $m=1$ and to
that with $x^{-2}$ (the centrifugal term, if applicable) in the remaining four
cases.

It has to be noted that for $\varepsilon=0$ and $\alpha=0$ (\ref{bhe})
reduces to the confluent hypergeometric differential equation.
Equation (\ref{emv}) also simplifies, which means that $m$ in that case is
restricted only to 1, 1/2 and 0, resulting in the Morse, harmonic oscillator
and Coulomb potentials \cite{jpa89}, respectively.
It is worth mentioning here that more
general potentials, the Natanzon confluent potentials \cite{natconf} can be
obtained by a more flexible choice of $z(x)$. In particular, it is possible
to identify combinations of several terms in (\ref{emv}) with the constant
($E$). As a result, in that case more complicated energy expressions are
obtained, and, at the same time, potential terms with fixed coupling
coefficients appear, originating from the Schwartzian derivative. The first
concrete example of Natanzon confluent potentials is the generalized
Coulomb potential \cite{jpa93,jmp98}, which bears the features of both the
Coulomb and the harmonic oscillator potentials that can also be obtained
from its special limits.

\subsection{The solutions}
\label{bhe-sum-2}

The general solution of the Schr\"odinger equation for the potentials
discussed here is written as
\begin{equation}
\psi(x)=z^{(\gamma-m)/2}(x)\exp\left(\frac{\delta}{2}z(x)
+\frac{\varepsilon}{4}z^2(x)\right)H_B(\gamma,\delta,\varepsilon;\alpha,q;z(x))\ ,
\label{sch-sol-bhe}
\end{equation}
where $H_B$ denotes the bi-confluent Heun function. Unfortunately, this
function is   far less well known than those solving other second-order
differential equations of mathematical physics, e.g. the (confluent)
hypergeometric function. Inspired by previous results
\cite{Ish_InvSqrt,Ish_CES_gen}, the  solutions of the bi-confluent Heun equation
(\ref{bhe}) can be sought for as an expansion in terms of Hermite
functions possessing shifted and scaled argument \cite{Ish17}:
\begin{equation}
H_{B}(z)=\sum_n c_n u_n(z)\ , \hspace{1cm}u_n(z)=H_{\alpha_0+n}(s_0(z+z_0))\ ,
\label{expan}
\end{equation}
where $\alpha_0$, $s_0$ and $z_0$ are complex constants. For integer values
of $\alpha_0$ the Hermite functions reduce to Hermite polynomials \cite{AS70}.

The Hermite functions satisfy the second-order differential equation
\begin{equation}
\frac{{\rm d}^2 u_n}{{\rm d}z^2}-2s_0^2(z+z_0)
\frac{{\rm d} u_n}{{\rm d}z}+2s_0^2(\alpha_0+n) u_n\ =0.
\label{hermf}
\end{equation}
As it has been shown in Ref. \cite{Ish17}, for specific choice of some
parameters, i.e. $s_0=\pm(-\varepsilon/2)^{1/2}$ and $z_0=\delta/\varepsilon$, a
three-term recurrence relation can be obtained for the coefficients $c_i$
in (\ref{expan}), after making use of the identities
\begin{eqnarray}
u'_n&=&2s_0(\alpha_0+n)u_{n-1}
\label{heq1}\ , \\
s_0(z+z_0)u_n&=&(\alpha_0+n)u_{n-1}+u_{n+1}/2\ .
\label{heq2}
\end{eqnarray}
This relation reads
\begin{eqnarray}
0&=&R_nc_n+Q_{n-1}c_{n-1}+P_{n-2}c_{n-2}
\label{rec1}\ , \\
R_n&=&\left(-\frac{2}{\varepsilon}\right)^{1/2}(\alpha_0+n)
[\alpha+(\alpha_0+n-\gamma)\varepsilon]
\label{rec2}\ , \\
Q_n&=&\mp\frac{1}{\varepsilon}
[\alpha\delta+(q+(\alpha_0+n)\delta)\varepsilon]
\label{rec3}\ , \\
P_n&=&(-2\varepsilon)^{-1/2}[\alpha+(\alpha_0+n)\varepsilon]\ ,
\label{rec4}
\end{eqnarray}
where the $\mp$ choice in (\ref{rec3}) corresponds to the choices
$s_0=\pm(-\varepsilon/2)^{1/2}$. The infinite summation in (\ref{expan})
reduces to a finite sum in certain situations specified in
Ref. \cite{Ish17}. For this, termination from both sides is required.
Termination from below occurs for $\alpha_0=0$, in which case the Hermite
functions reduce to Hermite polynomials, and also for
$\alpha_0=\gamma-\alpha/\varepsilon$. Termination from above is possible
if $c_{N+1}=0$ is prescribed, which implies that the accessory parameter
$q$ has to satisfy an $(N+1)$'th degree polynomial equation.

\section{The analysis of the {\it reduced} sextic oscillator in terms of
the BHE approach}
\label{comparison}

The formalism is now ready to establish how the general form of the sextic
oscillator presented in Ref. \cite{Ish17} is related to the one usually
discussed in the QES approach. For this, let
us make the $m=1/2$ choice in the differential equation (\ref{zdiff}),
and consider its solution with the natural choice $x_0=0$ and $\sigma=1$
discussed previously. Then we obtain $z(x)=x^2/4$, with the
substitution of which Eqs. (\ref{emv2}) and (\ref{sch-sol-bhe}) supply
the potential
\begin{eqnarray}
V(x)&=&\left(\gamma-\frac{1}{2}\right)\left(\gamma-\frac{3}{2}-1\right)x^{-2}
+\left(\frac{\delta^2}{16}-\frac{\alpha}{4}+\frac{\varepsilon}{8}(\gamma+1)\right)x^2
\nonumber\\
&&
+\frac{\delta\varepsilon}{32}x^4 +\frac{\varepsilon^2}{256}x^6\ ,
\label{vsextic}
\end{eqnarray}
the energy eigenvalues
\begin{equation}
E=-q-\frac{\gamma\delta}{2}\ ,
\label{esextic}
\end{equation}
and the corresponding bound-state wave functions
\begin{equation}
\psi(x)=(x^2)^{\frac{\gamma}{2}-\frac{1}{4}}\exp\left(\frac{\varepsilon}{64}x^4
+\frac{\delta}{8}x^2\right)H_B(\gamma,\delta,\varepsilon;\alpha,q;\frac{x^2}{4})\ .
\label{psisextic}
\end{equation}

On comparing (\ref{vsextic}) with (\ref{qes-sextosc}), as well as (\ref{psisextic})
with (\ref{bswf}), one finds that the two problems are equivalent, and the parameters
are related as
\begin{equation}
\gamma=2s\ ,\hspace{.4cm} \delta=-4b\ ,\hspace{.4cm} \varepsilon=-16a\ ,\hspace{.4cm}
\alpha=16aM .
\label{pars}
\end{equation}
With the $\delta=0$ ($b=0$) choice the quartic potential term is cancelled in
(\ref{vsextic}) leading to the {\it reduced} sextic oscillator discussed in
Section \ref{qes-sum}, while the exponential factor in the bound-state wave function
(\ref{psisextic}) is also simplified. Furthermore, it turns out that with the
$\alpha_0=0$ choice discussed in Subsection \ref{bhe-sum-2}, the bi-confluent
Heun function reduces to a finite sum of Hermite polynomials. The expansion
coefficients in the recurrence relation ({\ref{rec1}) also simplify:
\begin{eqnarray}
R_n&=&(-2\varepsilon)^{1/2} n(M-n+\gamma)\ ,
\label{rec2s} \\
Q_n&=&\mp q\ ,
\label{rec3s} \\
P_n&=&(-\varepsilon/2)^{1/2}(M-n)\ .
\label{rec4s}
\end{eqnarray}
Selecting the upper sign in (\ref{rec3s}) and with it, also the upper sign
in $s_0=\pm(-\varepsilon/2)^{1/2}$, as discussed in Subsection \ref{bhe-sum-2}
and in Ref. \cite{Ish17}, the bound-state wave functions can be expressed
as a finite sum of Hermite polynomials $H_n((-\varepsilon/32)^{1/2}x^2)$.

The termination of the series from above is secured by prescribing
$c_{N+1}=0$. Expressing this from (\ref{rec1}) in terms of $c_N$ and $c_{N-1}$
leads to an $(N+1)$'th degree polynomial in $q$. Since (\ref{esextic}) implies
$E=-q$, this relation determines the possible energy eigenvalues.
The integer parameter $M$ determines the number of energy levels to be discussed.
Taking $N=M$, the condition for the termination of the series is found to be
\begin{eqnarray}
M=0: && q=0\ ,
\label{m0}\\
M=1: && q^2+\gamma\varepsilon=0\ ,
\label{m1}\\
M=2: && q^3+2\varepsilon(2\gamma+1)q=0\ ,
\label{m2}\\
M=3: && q^4+10\varepsilon(\gamma+1)q^2+9\gamma(\gamma+2)\varepsilon^2=0
\label{m4}
\end{eqnarray}
It can be proven in a straightforward way that for these values of $M$ the
results of the QES approach outlined in Section \ref{qes-sum}, i.e.
Eqs. (\ref{m0wf})-(\ref{m3wf}) are reproduced. In particular,
Eqs. (\ref{m0})-(\ref{m4}) determine the expressions for $E_n$, while
the bound-state wave functions are constructed using the $c_i$ parameters
from the recurrence relation (\ref{rec1}) with (\ref{rec2s})-(\ref{rec4s}). In order to match these wave functions with those in
Eqs. (\ref{m0wf}) to (\ref{m3wf}), the explicit form of the Hermite polynomials
also has to be resolved.
Figures \ref{potfig} and \ref{solfig} display
the potential, the energy eigenvalues and the bound-state wave functions
for some parameters.

\begin{figure}[h]
\begin{center}
\includegraphics{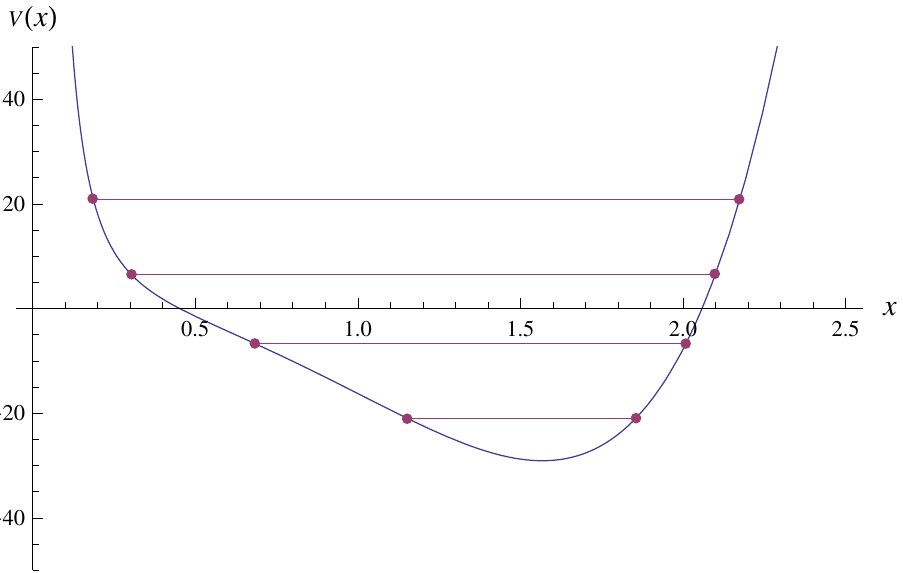}
\end{center}
\caption{The sextic potential (\ref{vsextic}) with $\gamma=2$, $\delta=0$,
$\varepsilon=-16$ and $\alpha/ \varepsilon=-3$ displayed together with
the energy eigenvalues located at $E_0=-20.926$, $E_1=-6.488$, $E_2=6.488$ and
$E_3=20.926$. }
\label{potfig}
\end{figure}

\begin{figure}[h]
\begin{center}
\includegraphics{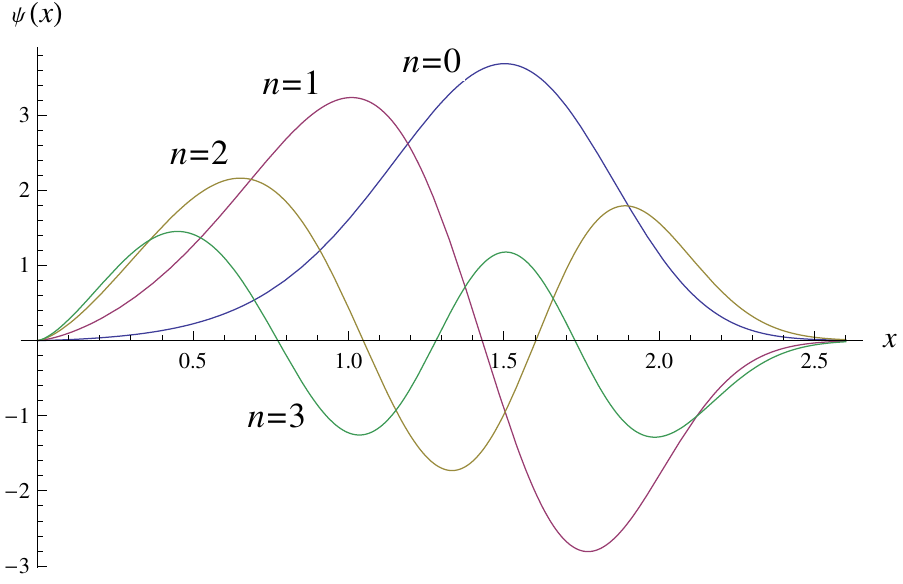}
\end{center}
\caption{Unnormalized  wavefunctions $\psi_n(x)$ plotted for $n=0$, 1, 2 and 3
corresponding to the four energy eigenvalues appearing in Fig. \ref{potfig}. }
\label{solfig}
\end{figure}

It is remarkable that the form of the wave functions (\ref{psisextic})
that is determined by the procedure summarized in Section \ref{bhe-sum}
and the choice of some parameters reproduces the known results obtained
through a completely different procedure, the QES approach. A technical
difference is that the polynomials of $x^2$ in the QES approach are
expanded in terms of Hermite polynomials here. Actually,
these polynomials have been discussed in detail from the mathematical
point of view in Ref. \cite{Ben96}, where they have been obtained as
the solutions of the {\it reduced} sextic oscillator with centrifugal
barrier, i.e. potential (\ref{qes-sextosc}) with $b=0$.

These results establish a
direct connection between the formalism of quasi-exactly solvable potentials
\cite{qes} and the method of expanding the solutions of the bi-confluent Heun
equation in terms of Hermite functions \cite{Ish17}.

\section{Summary and conclusions}
\label{summary}

In the present work we revisited the sextic oscillator potential problem and
presented its discussion from a new perspective. We demonstrated that this
problem can be solved by transforming the Schr\"odinger equation into the
bi-confluent Heun differential equation. The sextic oscillator is obtained
as one of the five potentials identified in Ref. \cite{Ish17}, corresponding
to a specific $z(x)$ transformation function in (\ref{zdiff}) with $m=1/2$.
It is also included implicitly in Ref. \cite{Batic1}, where the general
expression is discussed for potentials derived from the bi-confluent Heun
equation, as well as other versions of the Heun equation.
However, in these works the parametrization of the sextic oscillator was
general, i.e. it was different from the usual form discussed in the
quasi-exactly solvable framework, where the parameters are correlated.
The question whether the two approaches (i.e. the BHE and the QES) lead to
the same results was raised naturally.

Following the formalism of Ref. \cite{Ish17}, we sought for the bound-state
wave functions in terms of an expansion of Hermite functions with scaled and
shifted arguments. It turned out that the sextic oscillator can be obtained
after specifying some parameters of the model, in which case the Hermite
functions reduce to Hermite polynomials with shifted argument. Cancelling the
shift by selecting $\delta=0$, the {\it reduced} sextic oscillator is obtained,
i.e. the one with vanishing quartic term. The bound-state wave functions then
allow expansion in terms of Hermite polynomials with arguments involving $x^2$.
This expansion is reduced to a finite polynomial terminating with
$x^{2N}$ if the expansion coefficient of the $H_{N+1}$ Hermite polynomial
is forced to vanish: $c_{N+1}=0$. This condition implies an $(N+1)$'th
degree polynomial equation for the parameter $q$, the roots of which
supply the bound-state energy eigenvalues.

The explicit construction of the solutions up to $N=3$ recovers the results
obtained for the bound-state energy eigenvalues and wave functions derived
within the quasi-exactly-solvable formalism \cite{qes}. The bound-state
wave functions there contain polynomials of $x^2$, which are also obtained
from the condition of terminating a three-term recurrence relation. The
resulting polynomials have been analyzed from the mathematical point of view
in Ref. \cite{Ben96}, and were identified as a new set of orthogonal polynomials.

It is remarkable that the energy eigenvalues are located symmetrically
with respect to $E=0$. This seems to indicate the specific nature of the
{\it reduced} sextic oscillator in the sense that its parameters are
correlated and do not allow arbitrary independent values for $V_{-2}$,
$V_2$ and $V_6$ in (\ref{sextosc}).

The present investigation can be expanded to various directions. First, the
general version of the sextic oscillator can be considered with $\delta\ne 0$.
In this case the quartic potential term also appears, and the expansion of
the wave function contains Hermite functions with shifted arguments. One
can expect that forcing the termination of the series expansion also leads
to certain conditions and more complex energy spectrum and wave functions. The
energy eigenvalues and bound-state wave functions are less well-known in
the QES setting, so the comparison of the two approaches needs special care.
This will be done in a separate study.

It is also worthwhile to investigate the remaining four potentials identified
in Ref. \cite{Ish17}. The case with $m=0$, for example, yields the potential
with terms $x^p$, $p=-2$, $-1$, 1 and 2, i.e. a potential that contains
(shifted) harmonic oscillator, Coulombic and centrifugal terms alike. Special
versions of this potential have been studied before in terms of numerical
\cite{bessis} and expansion \cite{roych88} techniques.

The next level of complexity would be considering more general $z(x)$
transformation functions. One possibility is taking the one applied to
generate the generalized Coulomb potential \cite{jpa93,jmp98}, which belongs
to the Natanzon confluent class. It has both the harmonic oscillator
($z(x)\sim x^2$) and the Coulomb potential ($z(x)\sim x$) as a special limit,
reproducing the said two potentials there as limiting cases. The same two
limits appear in
the BHE approach too, corresponding to the sextic oscillator ($m=1/2$) and
the potential mentioned in the previous paragraph ($m=0$), offering an
interesting potential shape. From the technical point of view, this
potential would correspond to a specific problem contained implicitly in
the general discussion in Ref. \cite{Batic1}.

\section*{Acknowledgments}
This work was supported by the Hungarian Scientific Research
Fund -- OTKA (Grant No. K112962), the Armenian State Committee of Science
(SCS Grant No. 15T‐1C323), the Armenian National Science and Education Fund
(ANSEF Grant No. PS-4558) and the project “Leading Russian Research
Universities” (Grant No. FTI\_24\_2016 of the Tomsk Polytechnic University).
This research has been conducted within the scope of the International
Associated Laboratory IRMAS (CNRS-France and SCS-Armenia).

\end{document}